\UseRawInputEncoding
\PassOptionsToPackage{square,comma,numbers,compress}{natbib} 
\pdfoutput=1
\documentclass[preprint,12pt]{elsarticle}
\usepackage{natbib}
\usepackage{hyperref}
\usepackage{float}
\expandafter\let\csname equation*\endcsname\relax
\expandafter\let\csname endequation*\endcsname\relax
\let\eqref\relax
\expandafter\let\csname eqref\endcsname\relax
\usepackage{amsmath}
\usepackage{amssymb}
\usepackage{caption}
\usepackage{bookmark}
\usepackage{latexsym}
\usepackage{xcolor}
\usepackage{graphics}

\journal{Chinese Journal of Physics}

\begin{document}

\begin{frontmatter}

\title{A Family-nonuniversal $U(1)^\prime$ Model for Excited Beryllium Decays}

\author{Beyhan Puli\c{c}e$^*$}
\fntext[]{$*$ beyhanpulice@iyte.edu.tr}
\fntext[]{$*$ pulicebeyhan@gmail.com}

\address{Department of Physics, \.{I}zmir Institute of Technology, TR35430, \.{I}zmir, Turkey} 

\begin{abstract}
Excited beryllium has been observed to decay into electron-positron pairs with a $6.8~\sigma$ anomaly. The process is properly explained by a 17 MeV proto-phobic vector boson. In present work, we consider a family-nonuniversal $U(1)^{\prime}$ that is populated by a $U(1)^{\prime}$ gauge boson $Z^\prime$ and a scalar field $S$, charged under $U(1)^{\prime}$ and singlet under the Standard Model (SM) gauge symmetry. The SM chiral fermion and scalar fields are charged under $U(1)^{\prime}$ and we provide them to satisfy the anomaly-free conditions. The Cabibbo-Kobayashi-Maskawa (CKM) matrix is reproduced correctly by higher-dimension Yukawa interactions facilitated by $S$. The vector and axial-vector current couplings of the $Z^\prime$ boson to the first generation of fermions do satisfy all the bounds from the various experimental data. The $Z^\prime$ boson can have kinetic mixing with the hypercharge gauge boson and $S$ can directly couple to the SM-like Higgs field. The kinetic mixing of $Z^\prime$ with the hypercharge gauge boson, as we show by a detailed analysis, generates the observed beryllium anomaly. We find that beryllium anomaly can be properly explained by a MeV-scale sector with a minimal new field content. The minimal model we construct forms a framework in which various anomalous SM decays can be discussed.
\end{abstract}

\end{frontmatter}

\newpage
\section{Introduction}

The Atomki experiment has recently observed a $6.8\: \sigma$ anomaly \cite{Krasznahorkay2015} (see also \cite{Krasznahorkay2017-1,Krasznahorkay2017-2,Krasznahorkay2018}) in excited $^8Be$ nuclear decays, $^8Be^{^*} \rightarrow\: ^8Be\: e^+ e^-$, in both the distributions of the opening angles and the invariant masses of the electron-positron pairs (IPC). The SM predicts the angular correlation between the emitted $e^+ e^-$ pairs to drop rapidly with the separation angle. However, the experiment observed a bump with a high significance at a large angle of $\simeq 140^o$ which is consistent with the creation and subsequent decay of a new particle X with a mass $m_{X} = 16.7 \pm 0.35 (\text{stat}) \pm 0.5 (\text{sys})$ MeV. The best fit, which has $\chi^2/ dof = 1.07$, is obtained for the relative branching ratio $B(^8 Be^* \rightarrow {^8 Be} ~ X) / B(^8Be^* \rightarrow {^8 Be} ~\gamma) = 5.8 \times 10^{-6}$, assuming $Br(Z' \rightarrow e^-e^+ )=1$. In \cite{Krasznahorkay2019}, they observed also a peak in $e^- e^+$ angular correlations at $115^o$ with $7.2\: \sigma$ in  $21.01$ MeV $0^- \rightarrow 0^+$ transition of $^4 He$ and it is explained by a light particle with mass $m_x c^2 = 16.84 \pm 0.16 (\text{stat}) \pm 0.20 (\text{sys})$ MeV. It is likely the same particle with the one that is observed in \cite{Krasznahorkay2015}.

In recent interpretations of the experiment \cite{Feng2016-1,Feng2016-2}, possible particle physics interpretations of the $^8Be$ anomalous decays are examined and they concluded that a proto-phobic, spin-$1$ boson with a mass $\approx 17$ MeV fits the anomaly. They determine the bounds on the vector current couplings of the new gauge boson to the first generation of the SM fermions via a combination of the relevant experimental data. They propose two particle physics models, $U(1)_B$ and  $U(1)_{B-L}$ models, that are not initially anomaly-free therefore they add a new matter content to cancel the anomalies. Another recent interpretation \cite{Gu2016} makes an extension of the SM by two gauge groups, $U(1)_{Y^\prime} \times U(1)_X$, and they add a new matter content to get rid of the $Z-Z^{\prime}$ mass mixing.
In \cite{DelleRose2017}, they present a $U(1)^\prime$ extended $2$-Higgs doublet model for $^8Be$ anomalous decays. In \cite{Ellwanger2016} a pseudoscalar and in \cite{Kozaczuk2016} an axial vector candidate is presented. The extension of the minimal supersymmetric standard model (MSSM) by an extra $U(1)^\prime$ is discussed \cite{Demir2005} with $U(1)^\prime$ charges of the fields to be family-dependent and satisfy the anomaly-free conditions.

In this work, we extend the SM by a $U(1)^\prime$ which is populated by a light gauge boson $Z^\prime$  and a singlet scalar $S$. In the model, there are two mixings with the SM: the gauge kinetic mixing of the hypercharge gauge boson and the $Z^{^\prime}$ boson, and the quartic scalar mixing of the SM-like Higgs and the extra scalar. The masses of the gauge bosons are generated dynamically through the spontaneous symmetry breaking (SSB) via vacuum expectation values (vev) of the scalar fields. 

Our first intention in this work is to construct the framework of an anomaly-free, family-nonuniversal $U(1)^\prime$ model that fits the Atomki signal with a minimal field content. The model we present is able to explain the Atomki signal with a proto-phobic gauge boson with a mass of $\approx 17$ MeV. We find the couplings of the $Z^{^\prime}$ boson to the first generation of the SM fermions via the family-nonuniversal charges of the chiral fields that satisfy the anomaly-free conditions. We show that with these couplings we are able to explain the Atomki signal.

The paper is organized as follows. In Sec.\ref{sec-model}, we construct the framework of the family-nonuniversal $U(1)^\prime$ model. We summarize the experimental bounds in Sec.\ref{sec-bounds}. We give the vector and axial-vector current couplings of the $Z^{^\prime}$ boson to the first generation of the SM fermions in Sec.\ref{sec-Zprime-couplings}. We show that the CKM matrix is obtained properly in the model in Sec.\ref{CKM}.  In Sec.\ref{LHC-bound}, we consider the LHC bound on the decays of the SM Higgs. We summarize the model and discuss future prospects in Sec.\ref{sec-conclusion}.

\section{Family-nonuniversal $U(1)^\prime$ Model}
\label{sec-model}
In this section, we present the framework of the family-nonuniversal $U(1)^{\prime}$ model. We extend the SM gauge symmetry, $G_{SM}=SU(3)_{c} \times SU(2)_{L} \times U(1)_{Y}$, by an extra $U(1)^\prime$ symmetry 

\begin{align}
\label{new-symmetry}
G_{SM} \times U(1)^\prime.
\end{align}

The $U(1)^\prime$ quantum number assignment to chiral fermion and scalar fields is given in Tab.(\ref{table-charge-assignment}).

\begin{table}[ht!]
\begin{center}
    \begin{tabular}{| p{1.3cm} | p{1.3cm} | p{1.3cm} | p{1.3cm} |p{1.3cm} |}
    \hline
       & $SU(3)_c$ & $SU(2)_L$ & $U(1)_Y$ & $U(1)^\prime$ \\ \hline
       $Q_i$ & $3$ & $2$ & $1/6$ & $Q_{Q_i}$ \\ \hline
       $u_{R_i}$ & $3$ & $1$ &$2/3$ & $Q_{u_{Ri}}$\\ \hline
       $d_{R_i}$ & $3$& $1$ & $-1/3$ &$ Q_{d_{Ri}}$\\ \hline
       $L_i $& $1$ &$2$ & $-1/2$ & $Q_{L_i}$\\ \hline
       $e_{R_i}$ &$1$ & $1$ & $-1$ &$Q_{e_{R_i}}$\\ \hline
       $\hat{H}$ & $1$ & $2$ & $1/2$ & $Q_{H}$\\ \hline
       $\hat{S}$ & $1$ & $1$ & $0$ & $Q_{S}$\\ \hline
    \end{tabular}
    \caption{The gauge quantum numbers of the fields in the family-nonuniversal $U(1)^\prime$ model for $i=1,2,3$ that refers to the three generations of matter.}
\label{table-charge-assignment}    
\end{center}
\end{table}

\subsection{Mixing of Higgs Bosons}
 
The Lagrangian of the scalars in the family-nonuniversal $U(1)^{\prime}$ model is given by

\begin{align}
\mathcal{L}_{Higgs}&= \mathcal{L}^{SM}_{Higgs} + \mathcal{L}^{S}_{Higgs} + \mathcal{L}_{Higgs}^{mix}; \\
\mathcal{L}^{SM}_{Higgs} &= |\mathcal{D}_{\mu} \hat{H}|^{2}  + \mu^2 |\hat{H}|^{2}  - \lambda |\hat{H}|^{4}, \\
\mathcal{L}^{S}_{Higgs} &= |\mathcal{D}_{\mu} \hat{S}|^{2} + \mu_{s}^2 |\hat{S}|^{2} - \lambda_{s} |\hat{S}|^{4}, \\
\mathcal{L}_{Higgs}^{mix} &=  - \kappa |\hat{H}|^{2} |\hat{S}|^{2}
\end{align}
where the last equation contains a mixing term with a scalar mixing parameter $\kappa$. The hatted fields are used since we will use the fields without a hat in the mass-basis. 

We parametrize the SM-like Higgs $\hat{H}$ and the extra scalar $\hat{S}$ respectively as
 
\begin{align}
\label{Higgs-fields-Feynman-gauge}
   \hat{H}= \frac{1}{\sqrt{2}}\left( {\begin{array}{cc}
       \phi_{1} + i \phi_{2} \\
       v+\hat{h}+ i \phi_{3} 
     \end{array} } \right)  , 
     \: \: \: \: \: \:  \hat{S}=  \frac{1}{\sqrt{2}}\left( {\begin{array}{c}
         v_{s} + \hat{s} + i \phi_{s}  \end{array} } \right)
\end{align}
where $\phi_{1},\phi_{2},\phi_{3}$ and $ \phi_{s}$ are the Goldstone bosons; $v$ and $v_{s}$ are the vevs of the scalar fields that are real and positive. 

The scalar potential is bounded from below provided that

\begin{align}
\label{stability-pot}
\lambda > 0, ~~~~~\lambda_s > 0~~~~ \mathtt{and} ~~~~
 4 \lambda \lambda_s - \kappa^2 > 0.
\end{align}

For both nonvanishing values of the vevs, the minimum of the potential occurs at 

\begin{align}
\frac{v^2}{2} = \frac{2 \lambda_{s} \mu^2 - \kappa \mu_{s}^{2}}{4 \lambda \lambda_{s} -\kappa^2 },
\end{align} 

\begin{align}
 \frac{v_{s}^2}{2} = \frac{2 \lambda \mu_{s}^2 - \kappa \mu^{2}}{4 \lambda \lambda_{s} -\kappa^2}.
\end{align} 

These solutions are physical for $v^2 > 0$ and $v_s^2 > 0$ which lead to $\lambda_S \mu^2 > \kappa \mu_S^2 / 2$ and $\lambda \mu_S^2 > \kappa \mu^2 / 2$ if Eq.(\ref{stability-pot}) is satisfied. One can realize that for both nonvanishing vevs there are solutions for 

\begin{itemize}

\item  $\mu^2, \mu_s^2 > 0$ for both signs of $\kappa$,

\item  $(\mu^2 > 0, \mu_s^2 < 0)$ or $(\mu^2 < 0, \mu_s^2 > 0)$ for only $\kappa < 0$.

\item  There are not any solutions for $\mu^2, \mu_s^2 < 0$.
\end{itemize}
The scalar mass Lagrangian is given by

\begin{align}
\mathcal{L}_{scalar}^{mass} =  -V_{scalar} = -\frac{1}{2} \left( {\begin{matrix}{}
        \hat{h}&\hat{s} 
        \end{matrix} } \right)
        \left( {\begin{array}{cc}
              2 \lambda v^2 & \kappa v v_{s} \\
              \kappa v v_{s} &  2 \lambda_s v_{s}^2
            \end{array} } \right) 
\left( {\begin{array}{cc}
      \hat{h} \\
      \hat{s} 
    \end{array} } \right). 
\end{align}
The mass eigenstates, $(h,s)$, are obtained via the following transformation

\begin{align}
\label{h-s-mass-trans}
\left( {\begin{array}{cc}
      \hat{h} \\
      \hat{s} 
    \end{array} } \right) = 
    \left( {\begin{array}{cc}
                   \cos \alpha & \sin \alpha \\
                   -\sin \alpha & \ \cos \alpha
                  \end{array} } \right) 
    \left( {\begin{array}{cc}
                        h \\
                        s 
                      \end{array} } \right)
\end{align}
where the mixing angle is given by 

\begin{align}
\label{tan-2alpha}
\tan 2 \alpha = - \frac{\kappa v v_{s}}{\lambda v^{2} - \lambda_{s} v_{s}^{2}}.
\end{align}

The masses of the SM-like Higgs $h$ and the extra scalar $s$ are given by 

\begin{align}
\label{scalar-masses}
m_{h,s}^{2} = \lambda v^{2} + \lambda_{s} v_{s}^{2} \pm \sqrt{(\lambda v^{2} - \lambda_{s} v_{s}^{2})^{2} + \kappa^{2} v^{2} v_{s}^{2}}
\end{align}
where $\lambda v^2 > \lambda_s v_s^2$. In the limit of no scalar mixing, $\kappa \rightarrow 0$, the masses of the scalars in Eq.(\ref{scalar-masses}) reduce to 

\begin{align}
m_{h^0}^2 = 2 \lambda v^2 , \: \: \: m_{s^0}^2 = 2 \lambda_s v_s^2.
\end{align} 

\subsection{Mixing of Gauge Bosons}
\label{subsec-gauge-bosons}

The $U(1)^\prime$ couples to the SM hypercharge $U(1)_{Y}$ through kinetic mixing which leads to the most general gauge Lagrangian of $U(1)_{Y} \times U(1)^\prime$ 
 
\begin{align}
\label{gauge-Lag}
\mathcal{L}_{gauge}&= \mathcal{L}_{gauge}^{SM} + \mathcal{L}_{gauge}^{Z^{^\prime}}+\mathcal{L}_{gauge}^{mix}; \\ 
\mathcal{L}_{gauge}^{SM} &= -\frac{1}{4} \hat{B}_{\mu \nu} \hat{B}^{\mu \nu},  \\ 
 \mathcal{L}_{gauge}^{Z^\prime} &= -\frac{1}{4} \hat{Z^\prime}_{ \mu \nu} \hat{Z}^{\prime \mu \nu},  \\ 
 \mathcal{L}_{gauge}^{mix} &= -\frac{1}{2} \sin{\chi} \ \hat{B}_{\mu \nu} \hat{Z}^{\prime \mu \nu} 
\end{align}
where $\hat{B}_{\mu \nu}$ and $\hat{Z}^{\prime}_{\mu \nu}$ are the field strength tensors of $U(1)_{Y}$ and $U(1)^\prime$, respectively. The last equation contains a mixing term with a gauge kinetic mixing parameter $\chi$.

We diagonalize the field strength terms via a $GL(2,R)$ transformation 

\begin{eqnarray}
\label{gauge-trans}
\left(  {\begin{array}{c}
      \tilde{Z}^{\prime}_{\mu} \\
      \tilde{B}_{\mu} 
   \end{array} } \right)  
   = \left(  {\begin{array}{cc}
     \sqrt{1- \sin^2 \chi} & 0 \\
       \sin \chi & 1
       \end{array} } \right) 
     \left(  {\begin{array}{c}
          \hat{Z}^{\prime}_{\mu} \\
           \hat{B}_{\mu}
             
            \end{array} } 
            \right) 
\end{eqnarray}

where $\tilde{Z}^{\prime}_{\mu}$ and  $\tilde{B}_{\mu}$ are not the mass eigenstates yet.

In this basis, the general covariant derivative is given by

\begin{align}
\label{covariant-general}
\mathcal{D}_{\mu} = \partial_{\mu} + i g T^{i} W_{\mu}^{i} + i g^{\prime} Q_{Y} \tilde{B}_{\mu} +
i (e \tilde{g} Q^{\prime} + \eta g^{\prime} Q_{Y}) \tilde{Z}^{\prime}_{\mu}
\end{align}
where $T^{i}= \frac{1}{2} \sigma^{i}$ is  the third component of isospin in which $\sigma^{i}$ are the Pauli spin matrices with $i=1,2,3$; $W_{\mu}$ is the $SU(2)_L$ gauge field; $g$ and $g^\prime$ are the $SU(2)_L$ and $U(1)_Y$ gauge couplings, respectively. 

In Eq.(\ref{covariant-general}), we have introduced 

\begin{align}
\label{tan-chi}
\tilde{g} \equiv \frac{\hat{g}}{\cos \chi},~~ \: \: \eta \equiv - \tan \chi
\end{align}

where $\hat{g}$ is the normalized $U(1)^{\prime}$ gauge coupling 

\begin{align}
\hat{g} \equiv \frac{g_{U(1)^\prime}}{e}.
\end{align}
 
The mass squared matrix of the gauge bosons in the $(\tilde{B}_{\mu},\tilde{Z}^{\prime}_{\mu})$ gauge-basis is given by

\begin{align}
\label{gauge-mass-Lag} 
\mathcal{L}_{gauge}^{mass} &= \frac{1}{2} \left( {\begin{array}{ccc}
      \tilde{B}^{\mu} & W^{3 \mu}& \tilde{Z}^{^\prime \mu}
      \end{array} } \right) \nonumber \\
&. \left( {\begin{array}{ccc}
                   \frac{1}{4} v^2 g^{\prime 2} & -\frac{1}{4} v^2 g g^{\prime}& \frac{1}{2} g^{\prime} v^2 ( \frac{g^{\prime} \eta}{2} + e \tilde{g} Q_H) \\
                    -\frac{1}{4} v^2 g g^{\prime} & \frac{1}{4} v^2 g^{2} & -\frac{1}{2} g v^2 ( \frac{g^{\prime} \eta}{2} + e \tilde{g} Q_H) \\
                   \frac{1}{2} g^{\prime} v^2 ( \frac{g^{\prime} \eta}{2} + e \tilde{g} Q_H) &-\frac{1}{2} g v^2 ( \frac{g^{\prime} \eta}{2} + e \tilde{g} Q_H) &   v^2(\frac{ g^{\prime} \eta}{2} + e \tilde{g} Q_H)^2 + Q_{S}^{2} v_{s}^{2} e^2 \tilde{g}^{2}
                    \end{array} } \right) \nonumber \\
&. \left( {\begin{array}{ccc}
                        \tilde{B}_{\mu} \\
                        W^{3}_{\mu} \\
                        \tilde{Z}^{\prime}_{\mu}
                       \end{array} } \right). 
\end{align}

The mass eigenstates of the neutral gauge bosons are obtained via the following transformation

\begin{align}
\label{gauge-mass-eigenstate-trans}
\left( {\begin{array}{ccc}
                        \tilde{B}_{\mu} \\
                        W^{3}_{\mu} \\
                        \tilde{Z}^{^\prime}_{\mu}
                  \end{array} } \right) =
\left( {\begin{array}{ccc}
                        \cos \theta_{\scriptscriptstyle{W}} & -\sin \theta_{\scriptscriptstyle{W}} \cos \varphi & \sin \theta_{\scriptscriptstyle{W}} \sin \varphi \\
                        \sin \theta_{\scriptscriptstyle{W}} & \cos \theta_{\scriptscriptstyle{W}} \cos \varphi & -\cos \theta_{\scriptscriptstyle{W}} \sin \varphi  \\
                        0 & \sin \varphi & \cos \varphi
                       \end{array} } \right) 
\left( {\begin{array}{ccc}
                        A_{\mu} \\
                        Z_{\mu} \\
                        Z^{^{\prime}}_{\mu}
                  \end{array} } \right)                       
\end{align}
where $\theta_{\scriptscriptstyle{W}}$ is the Weinberg angle and $\varphi$ is the gauge mixing angle which is given by

\begin{align}
\label{gauge-mixing-angle}
\tan 2 \varphi = \frac{2 (g^{\prime} \eta + 2 e \tilde{g} Q_H ) \sqrt{g^{2} + g^{\prime 2}}}{(g^{\prime} \eta + 2 e \tilde{g} Q_H)^{2} + 4(\frac{v_{s}}{v})^{2} Q_{s}^{2} e^2 \tilde{g}^{2} - g^{2}-g^{\prime 2}}.  
\end{align}  

The masses of the physical gauge bosons read as

\begin{align}
M_A &= 0, \nonumber \\
M_{Z,Z^\prime}^2 &= \frac{1}{2} \Bigg \{M_{Z^0}^2 + M_{Z^{\prime 0 }}^2 + \Delta^2 
 \pm \sqrt{( M_{Z^0}^2 - M_{Z^{\prime 0}}^2 - \Delta^2  )^2 + 4 M_{Z^\prime{^0}}^2 \Delta^2} \Bigg \}
\end{align}
where 

\begin{align}
M_{Z^0}^2 &= \frac{1}{4} (g^2 + g^{\prime 2})v^2, \nonumber \\
M_{Z^{\prime 0}}^2 &= e^2 \tilde{g}^2 Q_{S}^2 v_{s}^2, \nonumber \\
\Delta &= v (\frac{g^\prime \eta}{2} + e \tilde{g} Q_H).
\end{align}

It is clear that if we impose the condition

\begin{align}
\label{no-gauge-mass-mixing-condition}
\Big( \frac{g^{\prime} \eta}{2} + e \tilde{g} Q_H \Big) = 0,
\end{align}
the gauge mixing angle in Eq.(\ref{gauge-mixing-angle}) vanishes identically. This ensures zero mixing between the $Z$ and $Z^{^\prime}$ so that the $Z^{^\prime}$ mass is set by the vev $v_s$ of the extra scalar 

\begin{align}
\label{Zprime-mass}
 M^2_{Z^\prime} = e^2 \tilde{g}^2 Q_S^2 v_s^2 .  
\end{align} 

The condition in Eq.(\ref{no-gauge-mass-mixing-condition}) can be relaxed. We know that the mixing of the $Z$ and $Z^{^\prime}$ can be at most at the level of the $Z^{^\prime}$ mass

\begin{align}
 \frac{1}{2} g^\prime v^2 \Big( \frac{g^{\prime} \eta}{2} + e \tilde{g} Q_H \Big) \lesssim M^2_{Z^\prime}
\end{align}
which gives

\begin{align}
\label{no-gauge-mass-mixing-extension}
\Big( \frac{g^{\prime} \eta}{2} + e \tilde{g} Q_H \Big) \lesssim 10^{-8}
\end{align}
for a $Z^\prime$ mass of $M_{Z^\prime} = 17$ MeV which implies $\tan 2 \varphi \lesssim 10^{-8}$. The current limit on the $Z-Z^\prime$ mixing angle from the LEP data is about $|\varphi|=10^{-3}-10^{-4}$ \cite{Erler2009}. It is thus clear that the $Z-Z^{\prime}$ mixing angle in our family-nonuniversal $U(1)^\prime$ model is well below the limit of the electroweak precision data. 

\subsection{Leptons and Quarks}

The kinetic Lagrangian of the fermions is given by
 
\begin{align}
\mathcal{L}_{fermion}^{kinetic} &= i\bar{Q}_{i} \gamma^{\mu} \mathcal{D}_{\mu} Q_{i} + i\bar{u}_{Ri} \gamma^{\mu} \mathcal{D}_{\mu} u_{Ri} + i\bar{d}_{Ri} \gamma^{\mu} \mathcal{D}_{\mu} d_{Ri} \nonumber \\
&+i\bar{L}_{i} \gamma^{\mu} \mathcal{D}_{\mu} L_{i} + i\bar{e}_{Ri} \gamma^{\mu} \mathcal{D}_{\mu} e_{Ri}
\end{align}
where $i=1,2,3$ is the family index, $Q_i$ is for the left-handed quark doublets and $(u_{Ri},d_{Ri})$ are for the right-handed quark singlets

\begin{align}
Q= \left( {\begin{array}{cc}
                     u_{Li} \\
                     d_{Li}
           \end{array} } \right) \: ,  u_{Ri}, d_{Ri},
\end{align}
and $L$ is for the left-handed lepton doublet and $e_{Ri}$ is for the right-handed lepton singlet

\begin{align}
L= \left( {\begin{array}{cc}
                     \nu_{Li} \\
                     e_{Li}
           \end{array} } \right) \: ,  e_{Ri}.
\end{align}

The Yukawa Lagrangian is 

\begin{align}
\label{yukawa-lag}
\mathcal{L}_{fermion}^{Yukawa} = -Y_{u} \bar{Q} \tilde{\hat{H}} u_{R} - Y_{d} \bar{Q} \hat{H} d_{R} - Y_{e} \bar{L} \hat{H} e_{R} + h. c.
\end{align}
where $(Y_{u},Y_{d},Y_{e})$ are the Yukawa matrices and $\tilde{\hat{H}}=i \sigma_{2} \hat{H}^{*}$. The gauge invariance conditions from the diagonal elements of the Yukawa interactions in Eq.(\ref{yukawa-lag}) are given by

\begin{align}
\label{gauge-invariance-conds}
Q_{u_{R_i}} &= Q_{Q_i} +Q_H, \nonumber \\
Q_{d_{R_i}} &= Q_{Q_i} -Q_H, \nonumber \\
Q_{e_{R_i}} &= Q_{L_i} -Q_H. \nonumber \\
\end{align}

It is clear that the conditions in Eq.$(\ref{gauge-invariance-conds})$ involve only the diagonal elements of the Yukawa interactions. 
They are actually general enough to cover also conditions coming from off-diagonal Yukawa entries. One will realize in Sec.(\ref{sec-Zprime-couplings}) that the $U(1)^\prime$ charges give rise to a specific mass matrix structure. The first two families of the up and down-type quarks have the same $U(1)^\prime$ charges while the third family has a different charge, which implies that $(M_u)_{13},(M_u)_{31},
(M_u)_{23},(M_u)_{32}$ and $(M_d)_{13}, \\ (M_d)_{31},
(M_d)_{23},(M_d)_{32}$ all vanish. These zeroes leave no Yukawa interactions between the first two families and the third family of the up and down-type quarks. There can arise thus no non-trivial gauge invariance conditions in these sectors. The general Yukawa interactions between the first two families are trivial in that their $U(1)^\prime$ charges are universal. Moreover, leptons have family-universal $U(1)^\prime$ charges. It, therefore, is clear that Eq.$(\ref{gauge-invariance-conds})$ covers all cases.

\section{Constraints from Experiments } 
\label{sec-bounds}
It is argued that the new boson is likely a vector boson  \cite{Feng2016-1,Feng2016-2} that couples to the SM fermion currents as

\begin{align}
\mathcal{L} \supset i Z^\prime_{\mu} J^{\mu} =  i  Z^\prime_{\mu} \sum_{i=u,d,e,\nu_e,...} \varepsilon^v_i e J_i^\mu \: , \: \: J_i^\mu = \bar{f}_i \gamma^\mu f_i
\end{align}
where $\varepsilon^v$ is the vector current couplings of the $Z^\prime$ with a superscript '$v$' referring to 'vector'. It is found that the vector current couplings of the $Z^\prime$ to the SM fermions are constrained from various experiments \cite{Feng2016-1,Feng2016-2}. The Atomki signal \cite{Krasznahorkay2015}, the neutral pion decay, $\Pi^0 \rightarrow X \gamma$, by NA$48/2$ experiment \cite{Batley2015,Raggi2016}, the SLAC E$141$ experiment \cite{Riordan1987,Bjorken2009,Essig2013}, constraint via the electron anomalous magnetic dipole moment $(g-2)_e$ \cite{Davoudiasl2014} and the $\bar{\nu}_e - e$ scattering by TEXONO \cite{Deniz2010} put the following constraints on the vector current couplings of the $Z^\prime$ to the first generation of the SM fermions 

\begin{align}
\label{coupling-constraints}
|\varepsilon^v_p| &\lesssim 1.2 \times 10^{-3}, \nonumber \\
|\varepsilon^v_n| &= (2-10) \times 10^{-3}, \nonumber \\
|\varepsilon^v_e| &= (0.2-1.4) \times 10^{-3}, \nonumber \\
\sqrt{\varepsilon^v_e \varepsilon^v_{\nu_e}} &\lesssim 7 \times 10^{-5}.
\end{align} 
The constraints on the couplings of the $Z^\prime$ from the neutral pion decay \cite{Feng2016-1,Feng2016-2} lead the $Z^\prime$ to be proto-phobic such that it has a suppressed coupling to the proton compared to the neutron 

\begin{align}
\label{protophobicity}
-0.067 < \frac{\varepsilon^v_p}{\varepsilon^v_n} < 0.078
\end{align}
where the nucleon couplings are explicitly given by

\begin{align}
\varepsilon^v_p = 2 \varepsilon^v_u + \varepsilon^v_d, \nonumber \\
\varepsilon^v_n = \varepsilon^v_u + 2 \varepsilon^v_d.
\end{align}

\section{$Z^{^\prime}$ Couplings}
\label{sec-Zprime-couplings}
In this section, we find the vector and axial-vector current couplings of the $Z^\prime$ that are able to explain
the Atomki anomaly. 
First, we show the vector and axial-vector current couplings of the $Z^\prime$ to the first generation of the fermions in terms of the model parameters including the $U(1)^\prime$ charges of the
related chiral fermions in Tab.$(\ref{table-Zprime-couplings-1st-solutions})$. 

\begin{table*}[h!]
\begin{center}
    \begin{tabular}{| p{7cm} | p{6cm} | }
    \hline
    $\varepsilon_u^v = \frac{1}{2} \epsilon + \frac{2}{3} \delta + \cos \varphi \tilde{g} \left ( \frac{Q_{Q_1}+Q_{u_{R_1}}}{2}\right ) $ & $\varepsilon_u^a = \frac{1}{2} \epsilon + \cos \varphi \tilde{g} \left ( \frac{Q_{Q_1} - Q_{u_{R_1}}}{2}\right )$ \\ \hline
     $\varepsilon_d^v = - \frac{1}{2} \epsilon - \frac{1}{3} \delta + \cos \varphi \tilde{g} \left ( \frac{Q_{Q_1}+Q_{d_{R_1}}}{2}\right ) $ & $\varepsilon_d^a = - \frac{1}{2} \epsilon + \cos \varphi \tilde{g} \left ( \frac{Q_{Q_1} - Q_{d_{R_1}}}{2}\right )$ \\ \hline
      $\varepsilon_e^v =  - \frac{1}{2} \epsilon - \delta + \cos \varphi \tilde{g} \left ( \frac{Q_{L_1}+Q_{e_{R_1}}}{2}\right )$ & $\varepsilon_e^a = - \frac{1}{2} \epsilon + \cos \varphi \tilde{g} \left ( \frac{Q_{L_1} - Q_{e_{R_1}}}{2}\right )$  \\ \hline       
      $\varepsilon_{\nu_e}^v = \frac{1}{2} \epsilon + \cos \varphi \tilde{g} \frac{Q_{L_1}}{2} $ & $\varepsilon_{\nu_e}^a =  \frac{1}{2} \epsilon + \cos \varphi \tilde{g} \left ( \frac{Q_{L_1}}{2}\right )$   \\ \hline
       \end{tabular}
\end{center}
\caption{The $Z^\prime$ couplings to the first generation of fermions in terms of the model parameters including the $U(1)^\prime$ charges of the related chiral fermions.}
\label{table-Zprime-couplings-1st-solutions}
\end{table*}

In Tab.(\ref{table-Zprime-couplings-1st-solutions}), we have introduced

\begin{align}
\label{epsilon-delta}
\epsilon &\equiv -\frac{1}{2} \Bigg ( (\cot \theta_{\scriptscriptstyle{W}} + \tan \theta_{\scriptscriptstyle{W}}) \sin \varphi + \frac{\cos \varphi}{\cos \theta_{\scriptscriptstyle{W}}} \eta   \Bigg ),  \\
\delta &\equiv \tan \theta_{\scriptscriptstyle{W}} \sin \varphi + \frac{\cos \varphi}{\cos \theta_{\scriptscriptstyle{W}}} \eta.
\end{align}

The SM chiral fermion and scalar fields are charged under $U(1)^\prime$. We determine the couplings by providing that the charges satisfy the anomaly-free conditions and the gauge invariance conditions. In order to avoid gauge and gravitational anomalies, the $U(1)^\prime$ charges of the chiral fields must satisfy the following conditions

\begin{eqnarray}
\label{anomaly-free-conds}
U(1)^\prime - SU(3) - SU(3): ~~~~0&=&\sum_i (2 Q_{Q_i} - Q_{u_{R_i}} - Q_{d_{R_i}}), \nonumber \\
U(1)^\prime - SU(2) - SU(2): ~~~~0&=& \sum_i (3 Q_{Q_i} + Q_{L_i}),  \nonumber \\
U(1)^\prime - U(1)_Y - U(1)_Y: ~~~~0&=& \sum_i (\frac{1}{6} Q_{Q_i} -\frac{1}{3} Q_{d_{R_i}} - \frac{4}{3} Q_{u_{R_i}} + \frac{1}{2} Q_{L_i} - Q_{e_{R_i}}), \nonumber \\
U(1)^\prime - \text{graviton - graviton}: ~~~~0&=& \sum_i (6 Q_{Q_i} - 3 Q_{u_{R_i}} - 3 Q_{d_{R_i}} + 2 Q_{L_i} - Q_{e_{R_i}} ) , \nonumber \\
U(1)^\prime - U(1)^\prime - U(1)_Y: ~~~~0&=& \sum_i (Q_{Q_i}^2 + Q_{d_{R_i}}^2 - 2 Q_{u_{R_i}}^2 -  Q_{L_i}^2 + Q_{e_{R_i}}^2), \nonumber \\
U(1)^\prime-U(1)^\prime-U(1)^\prime: ~~~~0&=& \sum_i (6 Q_{Q_i}^3 - 3 Q_{d_{R_i}}^3 - 3 Q_{u_{R_i}}^3 + 2 Q_{L_i}^3 - Q_{e_{R_i}}^3 ).
\end{eqnarray}

There are $16$ charges and $6$ anomaly-free conditions with additional conditions from Yukawa interactions, and as we show in Tab.(\ref{table-charge-solutions-1}), one could express $12$ charges in terms of $4$ free charges 

\begin{align}
Q_H, ~Q_{Q_2}, ~Q_{Q_3} ~\textnormal{and} ~Q_{L_3}. \nonumber
\end{align}

\begin{table*}[ht!]
\begin{center}
    \begin{tabular}{| p{4.3cm} | p{4.4cm} | p{3.7cm} | }
    \hline
    $Q_{Q_1} = Q_H -Q_{Q_2}-Q_{Q_3}$ & $Q_{u_{R_1}} = 2 Q_H -Q_{Q_2} -Q_{Q_3}$ & $Q_{d_{R_1}} = -Q_{Q_2} - Q_{Q_3}$  \\ \hline
      & $Q_{u_{R_2}} = Q_{Q_2} + Q_H$ & $Q_{d_{R_2}} = Q_{Q_2} - Q_{H}$ \\ \hline
       & $Q_{u_{R_3}} = Q_{Q_3} + Q_H$ & $Q_{d_{R_3}} = Q_{Q_3} - Q_{H}$ \\ \hline       
      $Q_{L_1} = -Q_H$&  $Q_{e_{R_1}} = -2 Q_H$ & \\ \hline
       $Q_{L_2} = -2 Q_H -Q_{L_3}$& $Q_{e_{R_2}} = -3 Q_H - Q_{L_3}$ &  \\ \hline
        &$Q_{e_{R_3}} = Q_{L_3} -Q_H$ & \\ \hline
       \end{tabular}
\end{center}
\caption{The $U(1)^\prime$ charge solutions of the chiral SM fermions by the gauge invariance and anomaly-free conditions.}
\label{table-charge-solutions-1}
\end{table*}

We parametrize the vector current coupling of the $Z^\prime$ boson to the proton as

\begin{align}
\label{proton-coupling-parametrization}
\varepsilon_p^v = 2 \varepsilon_u^v + \varepsilon_d^v = \delta^\prime
\end{align}
where we introduce a parameter $\delta^\prime$ which obeys the following experimental bound
 
\begin{align}
|\delta^\prime| \lesssim 10^{-3}.
\end{align}

One gets, via Eq.(\ref{proton-coupling-parametrization}), the following expression for $\delta$

\begin{align}
\delta = \delta^\prime - \frac{1}{2} \epsilon + \cos \varphi \tilde{g} \left ( 3 Q_{Q_2} + 3Q_{Q_3} -\frac{7}{2} Q_H \right )
\end{align}
which together with the charge solutions in Tab.(\ref{table-charge-solutions-1}) lead to the couplings in Tab.(\ref{table-Zprime-couplings-2nd-solutions}) with a vanishing gauge mixing angle, $\cos \varphi \rightarrow 1$. We apply the zero $Z-Z^\prime$ mixing limit from now on. 

\begin{table*}[ht!]
\begin{center}
    \begin{tabular}{| p{7cm} | p{3.5cm} | }
    \hline
    $\varepsilon_u^v =  \frac{1}{6} \epsilon + \frac{2}{3} \delta^\prime +  \tilde{g} (Q_{Q_ 2}+Q_{Q_ 3} -\frac{5}{6} Q_H )$ & $\varepsilon_u^a = \frac{1}{2} \epsilon - \frac{1}{2}    \tilde{g} Q_H$ \\ \hline
     $\varepsilon_d^v =  -\frac{1}{3} \epsilon - \frac{1}{3} \delta^\prime -  \tilde{g} (2 Q_{Q_ 2} + 2 Q_{Q_ 3} -\frac{5}{3} Q_H )$ & $\varepsilon_d^a = -\frac{1}{2} \epsilon + \frac{1}{2}  \tilde{g} Q_H$ \\ \hline
      $\varepsilon_e^v =  - \delta^\prime -  \tilde{g} (3 Q_{Q_ 2} + 3 Q_{Q_ 3} -2 Q_H )$ & $\varepsilon_e^a = -\frac{1}{2} \epsilon + \frac{1}{2}    \tilde{g} Q_H$  \\ \hline       
      $\varepsilon_{\nu_e}^v = \frac{1}{2} \epsilon - \frac{1}{2}  \tilde{g} Q_H $ & $\varepsilon_{\nu_e}^a = \frac{1}{2} \epsilon - \frac{1}{2}    \tilde{g} Q_H$   \\ \hline
       \end{tabular}
\end{center}       
\caption{The $Z^\prime$ couplings after imposing the charge solutions in Tab.(\ref{table-charge-solutions-1}) and parametrization of the vector current coupling of the $Z^\prime$ boson to the proton $\varepsilon^v_p = 2 \varepsilon^v_u + \varepsilon^v_d  \equiv \delta^\prime$ with $|\delta^\prime| \lesssim 10^{-3} $. Consideration of other constraints reduces the couplings in this table to the couplings in Tab.(\ref{table-Zprime-couplings-final}).}
\label{table-Zprime-couplings-2nd-solutions}
\end{table*}

The Lagrangian of the axial-vector current interaction of the $Z^\prime$ is given by

\begin{align}
\mathcal{L} \supset   i  Z^{^\prime}_{\mu} \sum_{i=u,d,e,\nu_e} \varepsilon^a_i e  \bar{f}_i \gamma^\mu \gamma^5 f_i
\end{align}
where $\varepsilon^a$ is the axial-vector current coupling with a superscript '$a$' referring to 'axial-vector'. 

We obtain the solutions of the free charges, $Q_{Q_ 2}, Q_{Q_ 3}$, $Q_H$ and $Q_{L_3}$, as follows.

\begin{itemize}

\item In the limit of minimal flavor violation, there holds the relation $\varepsilon_s^v = \varepsilon_d^v$ by which we obtain the solution 

\begin{align}
\label{QQ3-solution}
Q_{Q_3} = Q_H -2 Q_{Q_2}.
\end{align}

\item Next, we parametrize the vector current coupling of the $Z^\prime$ to the neutron

\begin{align}
\label{parameter-epsilon-prime}
\varepsilon^v_n = \varepsilon^v_u + 2 \varepsilon^v_d \equiv \epsilon^\prime
\end{align}
where the parameter $\epsilon^\prime$ satisfies the following experimental constraint

\begin{align}
|\epsilon^\prime| \approx  (2-10) \times 10^{-3}.
\end{align}
Then, by Eq.(\ref{QQ3-solution}) and Eq.(\ref{parameter-epsilon-prime}), we obtain the solutions of $Q_{Q_2}$ and $Q_{Q_3}$ 

\begin{align}
\label{QQ2-QQ3-solutions}
Q_{Q_2} &= \frac{1}{3 \tilde{g}} \left ( \epsilon + \epsilon^\prime \right), \nonumber \\
Q_{Q_3} &= \frac{1}{3 \tilde{g}} \left ( \epsilon - 2 \epsilon^\prime \right).
\end{align}

\item  The axial-vector coupling to the electron vanishes, $\varepsilon_e^a =0$, identically via the zero $Z-Z^\prime$ mixing condition in Eq.(\ref{no-gauge-mass-mixing-condition}) as well as the axial-vector current couplings to the up and down quarks $\varepsilon^a_u = \varepsilon^a_d =0$; the vector and axial-vector current couplings to the electron neutrino, $\varepsilon^v_{\nu_e}=\varepsilon^a_{\nu_e}=0$ by the following $U(1)^\prime$ charge solution of the SM-like Higgs boson 

\begin{align}
\label{Higgs-U(1)prime-charge}
Q_H = \frac{\epsilon}{\tilde{g}}.
\end{align}

Using the solution of $Q_H$ in \ref{Higgs-U(1)prime-charge}, we get $\eta \lesssim 10^{-4}$ which well agrees with the bounds.

\item The axial-vector current coupling of the $Z^\prime$ boson to the electron is constrained from the neutral pion decay process, $\Pi^0 \rightarrow e^+ e^-$  \cite{Abouzaid2006}. The matrix element of this process is proportional to $\varepsilon^a_e (\varepsilon^a_u-\varepsilon^a_d)$ \cite{Kahn2007}. However, in our model the axial-vector current coupling of the $Z^\prime$ to the electron vanishes, $\varepsilon_e^a =0$, as well as the axial-vector current couplings to the up and down quarks $\varepsilon^a_u = \varepsilon^a_d =0$. Therefore there arise no constraints from this rare process. The axial-vector current coupling of the $Z^\prime$ to the electron is constrained also from the atomic parity violation \cite{Porsev2009} and the parity-violating M$\o$ller scattering \cite{Anthony2005} which constrain $\varepsilon^a_e \varepsilon^v_q$ and $\varepsilon^a_e \varepsilon^v_e$, respectively. It is obvious that due to vanishing $\varepsilon_e^a$, there arise no constraints from these processes.

As a result of these, the vector and axial-vector current couplings of the $Z^\prime$ to the first generation of the SM fermions take the forms in Tab.(\ref{table-Zprime-couplings-final}).

\begin{table}[h!]
\begin{center}
    \begin{tabular}{| p{3cm} | p{2cm} | }
    \hline
    $\varepsilon_u^v = \frac{2}{3} \delta^\prime-\frac{1}{3} \epsilon^\prime$ & $\varepsilon_u^a = 0$ \\ \hline
     $\varepsilon_d^v = -\frac{1}{3} \delta^\prime + \frac{2}{3} \epsilon^\prime$ & $\varepsilon_d^a = 0$ \\ \hline
      $\varepsilon_e^v = \epsilon^\prime- \delta^\prime$ & $\varepsilon_e^a =0$ \\ \hline       
      $\varepsilon_{\nu_e}^v = 0 $ & $\varepsilon_{\nu_e}^a = 0$   \\ \hline
       \end{tabular}
\end{center}
\caption{The $Z^\prime$ couplings to the first generation of the SM fermions that explain the Atomki signal by $\varepsilon^v_p = 2 \varepsilon^v_u + \varepsilon^v_d  \equiv \delta^\prime$, $|\delta^\prime| \lesssim  10^{-3}$ and $\varepsilon^v_n = \varepsilon^v_u + 2 \varepsilon^v_d \equiv \epsilon^\prime$, $|\epsilon^\prime| \approx (2-10) \times 10^{-3}$. The couplings of the $Z^{^\prime}$ are proto-phobic Eq.(\ref{protophobicity}), and satisfy the experimental constraints in Eq.(\ref{coupling-constraints}).}
\label{table-Zprime-couplings-final}
\end{table}

\item The $Z^\prime$ couplings in Tab.(\ref{table-Zprime-couplings-final}) satisfy all the experimental constraints that explain the Atomki signal, and due to the zero coupling to the neutrinos, we have $Br(Z' \rightarrow e^-e^+) = 1$.

\item 
In Tab.(\ref{table-Zprime-couplings-final}), we present the $Z^\prime$ couplings to the first generation of the SM fermions that explain the Atomki signal. The couplings of the $Z^{^\prime}$ are proto-phobic Eq.(\ref{protophobicity}), and satisfy the experimental constraints in Eq.(\ref{coupling-constraints}) by $\varepsilon^v_p = 2 \varepsilon^v_u + \varepsilon^v_d  \equiv \delta^\prime$, $|\delta^\prime| \lesssim  10^{-3}$ and $\varepsilon^v_n = \varepsilon^v_u + 2 \varepsilon^v_d \equiv \epsilon^\prime$, $|\epsilon^\prime| \approx (2-10) \times 10^{-3}$.

\item As one can realize, our model is proto-phobic in both the vector and axial-vector current interactions. The axial-vector current couplings to the up and down quarks vanish identically via the zero $Z-Z^\prime$ mixing condition in Eq.(\ref{no-gauge-mass-mixing-condition}), so the $Z^\prime$ has purely vector current interactions with the up and down quarks. 

\item The vector current coupling to the electron does not vanish as it should not for the IPC, and it can lie within the experimental range. The axial-vector current coupling to the electron vanishes identically via the zero $Z-Z^\prime$ mixing condition in Eq.(\ref{no-gauge-mass-mixing-condition}).

\item The experimental constraints require the vector current coupling to the electron neutrino to be significantly below the vector current coupling to the neutron. The vector and axial-vector current couplings to the electron neutrino vanish identically with zero $Z-Z^\prime$ mixing condition in Eq.(\ref{no-gauge-mass-mixing-condition}), and this obviously satisfies the experimental data. 

\item In order to have universal charges in the lepton sector, we assume 

\begin{align}
Q_{L_3} = - Q_H.
\end{align}

As a result of these, the first two families of the quarks have the same $U(1)^\prime$ charges which are different from the third family charge, and the leptons have universal $U(1)^\prime$ charges as we show in Tab.(\ref{table-charge-solutions-final}).

\begin{table*}[ht!]
\begin{center}
    \begin{tabular}{| p{4.4cm} | p{4.9cm} | p{4.9cm} |}
   \hline
    $Q_{Q_1}=Q_{Q_2}=\frac{1}{3 \tilde{g}} (\epsilon + \epsilon^\prime)$ & $Q_{u_{R_1}}=Q_{u_{R_2}}=\frac{1}{3 \tilde{g}} (4 \epsilon + \epsilon^\prime)$  & $Q_{d_{R_1}}=Q_{d_{R_2}}=\frac{1}{3 \tilde{g}} (-2 \epsilon + \epsilon^\prime)$  \\ \hline
    $Q_{Q_3}=\frac{1}{3 \tilde{g}} (\epsilon - 2 \epsilon^\prime)$ & $Q_{u_{R_3}}=\frac{2}{3 \tilde{g}} (2 \epsilon - \epsilon^\prime)$ & $Q_{d_{R_3}}=-\frac{2}{3 \tilde{g}} (\epsilon + \epsilon^\prime)$  \\ \hline       
      $Q_{L_1}=Q_{L_2}=Q_{L_3}=-\frac{\epsilon}{\tilde{g}}$ &  $Q_{e_{R_1}} = Q_{e_{R_2}}=Q_{e_{R_3}}=-\frac{2 \epsilon}{\tilde{g}}$ & \\ \hline
       \end{tabular}
\end{center}
\caption{The $U(1)^\prime$ charges of the chiral SM fermions. One obtains the $Z^\prime$ couplings in Tab.(\ref{table-Zprime-couplings-final}) if these charge solutions are put into the couplings in Tab.(\ref{table-Zprime-couplings-2nd-solutions}).}
\label{table-charge-solutions-final}
\end{table*}

\end{itemize}

\section{CKM Matrix}
\label{CKM}
 
There are several texture-specific quark mass matrices in the literature \cite{Rasin1998, Branco1999, Fritzsch2000, Xing2003, Branco2007, Gupta2011, Gupta2013}. The goal has always been avoiding a large number of parameters in these mass matrices. Some elements of these matrices are assumed to be zero and they are generally referred to as 'texture zero matrices'. These kinds of matrices provide a viable framework to obtain the flavor mixing matrix, the CKM matrix, which is compatible with the current data \cite{Patrignani2016}.  

For definiteness, we focus here on the texture-specific quark mass matrices in \cite{Fritzsch1997,Fritzsch1999}

\begin{align}
\label{Fritzsch-mass-matrices}
 M_{u,d} =
 \left( {\begin{array}{ccc}
        \times & \times  & 0 \\
        \times & \times & \times \\
         0 & \times &  \times
        \end{array} } \right)
\end{align}
which are known to reproduce the CKM matrix. The viability of these mass matrices are analyzed in \cite{Ahuja2016} by showing the compatibility with the CKM matrix.

In our model, the Higgs field leads to $(M_{u,d})_{13}=0, (M_{u,d})_{31}=0$ and $(M_{u,d})_{23}=0, (M_{u,d})_{32}=0$. In order to match to Eq.(\ref{Fritzsch-mass-matrices}), we need to induce matrix elements $(M_{u,d})_{23} \neq 0$ and $(M_{u,d})_{32} \neq 0$. One way to do this is by higher-dimensional operators \cite{Buchmuller1985, Barger2003, Grzadkowski2010, Murdock2010}. Then, as a minimal approach that fits to our $U(1)^\prime$ set up, we introduce the Yukawa interactions 

\begin{align}
\label{high-d-Yukawa}
\mathcal{L}  \supset \lambda_u^{23} \left(\frac{S}{\Lambda}\right)^{\delta_u^{23}} \bar{Q}_2 \tilde{\hat{H}} t_R
+ \lambda_u^{23} \left(\frac{S S^*}{\Lambda^2}\right)^{\delta_d^{23^\prime}} \left(\frac{S}{\Lambda}\right)^{\delta_d^{23}} \bar{Q}_2 \hat{H} b_R + h.c.
\end{align}  
where $\lambda_{u}^{23}$ is the Yukawa coupling, $\Lambda$ is the mass scale for flavor physics, $\delta_{u,d}^{23}$ and $\delta_{d}^{23^\prime}$ are parameters that will be determined below. We get the gauge invariance conditions by Eq.(\ref{high-d-Yukawa}) as follows

\begin{align}
-Q_{Q_2} - Q_H + Q_{u_{R_3}} + \delta_u^{23}.Q_S = 0, \\ \nonumber
-Q_{Q_2} + Q_H + Q_{d_{R_3}} + \delta_d^{23}.Q_S = 0
\end{align}
which lead to the following solutions

\begin{align}
\delta_u^{23} = \delta_d^{23} = \frac{\epsilon^\prime}{Q_S \tilde{g}}
\end{align}
by using the charge solutions in Tab.(\ref{table-charge-solutions-final}). This method of generating the hierarchy can be extended to the other Yukawa entries (in terms of their 33 entries or few other entries) \cite{Buchmuller1985, Barger2003, Grzadkowski2010, Murdock2010}. 

The parameters $\delta_u^{23}$ and $\delta_d^{23}$ are positive integers, so that we adopt $Q_S = \frac{\epsilon^\prime}{\tilde{g}}$ to obtain $\delta_u^{23} = \delta_d^{23} = 1$. This solution of $Q_S$ leads to $v_s \approx \mathcal{O}(10)$ GeV for a $17$ MeV $Z^\prime$ boson. The charge of the extra scalar $\hat{S}$ is $Q_S \approx \mathcal{O}(10^{-2})$ for the coupling $\tilde{g} \approx \mathcal{O}(10^{-1})$. If we use the optimized values of the matrix elements of $(M_{u,d})_{23}$ from \cite{Ahuja2016}, we find $\delta_d ^{23^\prime} \approx 2$ for $\Lambda \approx \mathcal{O}(10)$ GeV and $\lambda_{u,d}^{23} = 1$. 

The solutions via Eq.(\ref{high-d-Yukawa}) are not necessarily specific to the texture in Eq.(\ref{Fritzsch-mass-matrices}). One can consider different textures and generate the same CKM structure by modifications or extensions of Eq.(\ref{high-d-Yukawa}).

In the present model in the interaction basis, the couplings of the $Z^\prime$ to the SM quarks are diagonal but nonuniversal. This nonuniversality gives rise to flavor changing neutral currents (FCNCs). From $B^0 - \bar{B}^0$ mixing there arise stringent constraints for these FCNCs \cite{Becirevic2016,Kumar2019}

\begin{align}
|\epsilon^{L(R)}| \lesssim 10^{-6} 
\end{align}
where $\epsilon^{L(R)}$ is the chiral coupling of the $Z^\prime$ to the $\bar{s}\gamma^\mu b$ current. 

In the present model, the chiral couplings in the down quark sector are given by 

\begin{align}
g_{d_L} \equiv diag(g_{d_L}^1,g_{d_L}^1,g_{d_L}^3), \\
g_{d_R} \equiv diag(g_{d_R}^1,g_{d_R}^1,g_{d_R}^3)
\end{align}
where $ g_{d_L}^1 = g_{d_R}^1 = \frac{\epsilon^\prime}{3},
g_{d_L}^3 = g_{d_R}^3 = - \frac{2 \epsilon^\prime}{3}$. If we introduce the CKM matrix, the chiral couplings in the quark mass eigenstate basis become 

\begin{align}
\label{chiral-couplings}
\epsilon_{sb}^L &\equiv (V_{CKM}  g_{d_L} V_{CKM}^\dagger)_{23}, \\
\epsilon_{sb}^R &\equiv (V_{CKM}^\dagger  g_{d_R} V_{CKM})_{23}.
\end{align}
Then, one obtains the following condition from the chiral couplings 

\begin{align}
|\epsilon^\prime|=2 \times 10^{-3}.
\end{align}

\section{LHC Bound}
\label{LHC-bound}
In our family-nonuniversal $U(1)^\prime$ model, the SM-like Higgs boson is charged under $U(1)^\prime$ which leads the decay ($h \rightarrow Z^{^\prime} Z^{^\prime}$) that should be sufficiently small such that the branching fraction of the SM-like Higgs to the $Z^\prime$ boson pairs has to be $BR(h \rightarrow Z^{^\prime} Z^{^\prime}) \lesssim 10\%$
\cite{Curtin2013,Lee2013}.

The decay rate of this process is given by

\begin{align}
\Gamma (h \rightarrow Z^\prime Z^\prime) = \frac{3}{32 \pi m_h} \xi^2 \Bigg (1-\frac{4 M_{Z^\prime }^2}{m_h^2} \Bigg)^{1/2}
. \Bigg( 1-\frac{m_h^2}{3 M_{Z^\prime}^2} + \frac{m_h^4}{12 M_{Z^\prime}^4}\Bigg)
\end{align} 
where we have introduced

\begin{align}
\xi \equiv 4 \Bigg[ \cos\alpha \sin^2 \theta_W \eta^2 \frac{M_Z^2}{v} - \sin \alpha \frac{M_{Z^\prime}^2}{v_s} - \frac{\cos \alpha}{2 \cos \theta_W} v \Bigg ( g^\prime - \frac{e}{2 \cos \theta_W} \Bigg)  e \eta^2 \Bigg].
\end{align}

\begin{figure}[ht!]
\center
\includegraphics[width=0.7
\textwidth]{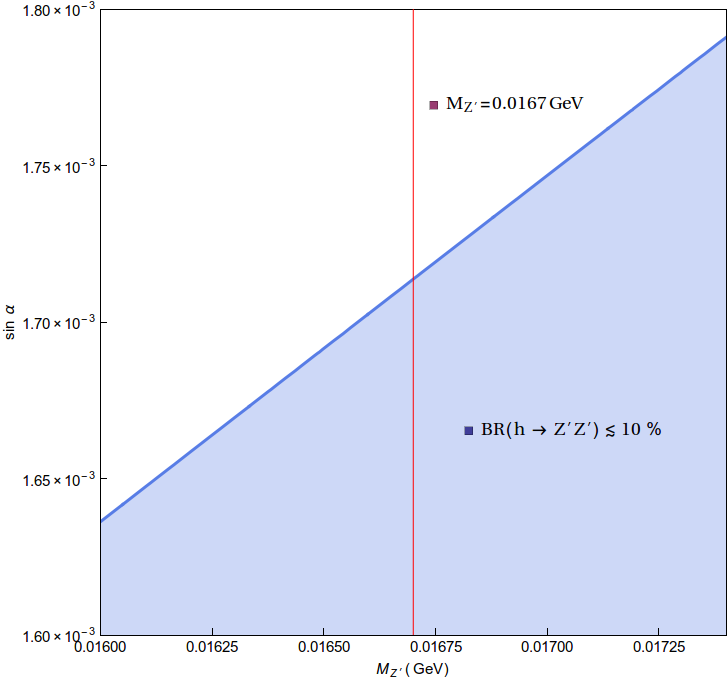}
\caption{We show the region where the partial decay width $\Gamma (h \rightarrow Z^\prime Z^\prime)$ is less than $10 \%$ of the SM Higgs total decay  width $BR(h \rightarrow Z^\prime Z^\prime) \lesssim 10\%$. The Higgs mixing angle is $\sin \alpha \sim \mathcal{O}(10^{-3})$ for $m_h=125.09$ GeV and $\eta = 10^{-4}$. The vertical red line is for the $Z^\prime$ boson mass $M_{Z^\prime}$ determined via the experimental data.}
\label{Bratio-higgs-ZpZp}
\end{figure}

In Fig.(\ref{Bratio-higgs-ZpZp}), we show the region where the partial decay width $\Gamma (h \rightarrow Z^\prime Z^\prime)$ is less than $10 \%$ of the SM Higgs total decay  width 

\begin{align}
BR(h \rightarrow Z^\prime Z^\prime) = \frac{\Gamma (h \rightarrow Z^\prime Z^\prime)}{\Gamma_{total}^{SM}(h)+\Gamma (h \rightarrow Z^\prime Z^\prime)} \lesssim 0.10
\end{align}
where $\Gamma_{total}^{SM}(h) = 4.07 \times 10^{-3}$ GeV \cite{Heinemeyer2013}.

The scalar mixing angle is found as $\sin \alpha \sim \mathcal{O}(10^{-3})$, and accordingly, the scalar mixing parameter becomes $\kappa \sim \mathcal{O}(10^{-3})$ which is required for  $BR(h \rightarrow Z^\prime Z^\prime) \lesssim 10\%$, for the SM Higgs boson mass of $m_h = 125.09$ \cite{Patrignani2016} and $\eta = 10^{-4}$. The scalar mixing remains at the same order for different values of the kinetic mixing $\eta = 10^{-5}, 10^{-6}$. 

The decay process ($h \rightarrow Z Z^\prime $) would also be relevant, however the $(h Z Z^{^\prime})$ vertex factor, which is given by

\begin{align}
h Z Z^\prime : -\frac{\cos \alpha}{\sin 2 \theta_W} v e \Big( \frac{g^{\prime} \eta}{2} + e \tilde{g} Q_H \Big),
\end{align}
is proportional to the left-hand side of the zero $Z-Z^\prime$ mixing condition in Eq.(\ref{no-gauge-mass-mixing-condition}). Therefore this vertex is zero, and there arise no constraints from this decay.

\section{Summary and Outlook}
\label{sec-conclusion}
In this work, we construct the framework of a family-nonuniversal $U(1)^\prime$ model, which is a minimal and an anomaly-free extension of the SM that is able to explain the $6.8 ~ \sigma$ anomaly in $^8Be$ nuclear decays at the Atomki pair spectrometer experiment. 

One possible interpretation of the Atomki signal is a spin-$1$, proto-phobic gauge boson with a mass of $ \approx 17$ MeV. We present a family-nonuniversal $U(1)^\prime$ model with its associated $Z^\prime$ boson with a mass of $\approx 17$ MeV which fulfills all the experimental constraints on its vector and axial-vector current couplings to the first generation of fermions that are necessary to explain the $^8Be$ anomalous decays. 

The previously proposed models have a large content of new fields. However, we have a minimal new field content with the $Z^\prime$ boson and the extra scalar. Our family-nonuniversal $U(1)^\prime$ model is an anomaly-free extension of the SM with a minimum field content that can explain the observed beryllium anomaly. 

The CKM matrix is reproduced correctly by higher-dimensional Yukawa interactions facilitated by $S$. The model provides new couplings to probe new physics at low energies. It may provide a framework for anomalous SM decays and forms a framework in which various low-energy phenomena can be addressed. Processes such as $s s^{*} \rightarrow f \bar{f}$ and $Z^\prime Z^\prime \rightarrow f \bar{f}$ might be effective around the Big Bang Nucleosynthesis (BBN) phase in accordance with the thermal equilibrium. BBN as a probe of the early universe puts constraints on physics beyond the SM, and the processes that are mentioned above might be relevant to study in the early universe. A singlet fermionic dark matter candidate can also be studied in the present framework such that it may interact with the SM via a scalar mediator. These astrophysical and cosmological implications can be relevant for future work. 

\section*{Acknowledgements}
The author thanks Durmu\c{s} Demir for suggestions and discussions on the problem.


\end{document}